\documentclass[a4paper,12pt]{article}

\usepackage{graphicx}
\usepackage{dcolumn}
\usepackage{bm}

\usepackage{amssymb,amsfonts,amsmath,epsfig,float,graphicx}
\usepackage{hyperref} 
\usepackage{xcolor}
\usepackage{ulem}

\usepackage[T1]{fontenc}

\usepackage{mathrsfs}

\newtheorem{theorem}{Theorem}[section]
\newtheorem{proposition}[theorem]{Proposition}
\newtheorem{lemma}[theorem]{Lemma}
\newtheorem{corollary}[theorem]{Corollary}

\def\Io{{\mathbb I}}

\def\Mo{{\mathbb M}}

\def\Ro{{\mathbb R}}

\def\beginproof{\par\strut\vskip 0.1cm\noindent{\bf Proof}\par}
\def\endproof{\par\strut\hfill$\square$\par\vskip 0.5cm}

\def\Io{{\mathbb I}}

\def\Mo{{\mathbb M}}
\def\Ro{{\mathbb R}}

\newcommand{\be}{\begin{equation}}
\newcommand{\ee}{\end{equation}}

\def\Tr{\mbox{ Tr }}

\def\tinyT{{\scriptscriptstyle T}}

\def\ihat{{\hat i}}
\def\jhat{{\hat j}}
\def\khat{{\hat k}}
\def\chat{{\hat c}}
\def\dhat{{\hat l}}
\def\hhat{{\hat n}}
\def\mhat{{\hat m}}
\def\phat{{\hat p}}
\def\qhat{{\hat q}}
\def\Splus{{S_\oplus}}
\def\Hplus{{H_\oplus}}

\def\Deltaplus{{\Delta_\oplus}}
\def\Deltaplusit{\Delta_\oplus^{\ihat t}}

\def\sep{\,:\,}
\def\antisep{\,\cdot\cdot\,\,}

\def\Kplus{{\mathscr K}_\oplus}
\def\Lplus{{\mathscr L}_\oplus}

\def\evec{{\mathbf e}}

\def\mvec{{\mathbf m}}
\def\xvec{x}
\def\yvec{y}
\def\zvec{z}
\def\uvec{u}
\def\vvec{v}
\def\wvec{w}

\begin{document}

\title{Pairs of Subspaces, Split Quaternions\\
                     and the Modular Operator}
\author{Jan Naudts\\
Universiteit Antwerpen\\
\small Physics Department, Universiteitsplein 1, 2610 Antwerpen, Belgium\\
\small	\url{Jan.Naudts@uantwerpen.be}\\[2ex]
%
Jun Zhang\\
University of Michigan\\
\small Ann Arbor, MI U.S.A.\\
\small \url{junz@umich.edu}\\
}

\date{}

\maketitle

\begin{abstract}
We revisit the work of Rieffel and van Daele on pairs of subspaces of a real Hilbert space, while relaxing
as much as possible the assumption that all the relevant subspaces are in general positions with respect to each other.
We work out, in detail, how two real projection operators lead to the construction of a complex Hilbert space
where the theory of the modular operator is applicable, with emphasis on
the relevance of a central extension of the group of split quaternions.
Two examples are given for which the subspaces have unequal dimension and therefore are not in generic position.

\end{abstract}


\section{Introduction}

Pairs of subspaces of a real Hilbert space $\mathscr V$ with inner product $s$
have been studied by Halmos \cite{HPR69}.
He proves that given two closed subspaces $\mathscr K$ and $\mathscr L$ of $\mathscr V$,
one can represent the orthogonal projections $P$ and $Q$ onto $\mathscr K$, respectively $\mathscr L$,
by 2-by-2 matrices with operator entries. This allows for a geometric analysis of the 
von Neumann algebra generated by $P$ and $Q$. See Theorem 2 of \cite{HPR69}
and the discussion which follows it.

Rieffel and van Daele \cite{RvD77} elaborate the link between Halmos' work and the theory
of the modular operator, also known as Tomita-Takesaki theory \cite{TM70, BR79}.
Given the two closed subspaces $\mathscr K$ and $\mathscr L$ there
exist a partial isometry $J$ and a positive operator $\Delta$ 
which are the analogues of, respectively, the modular conjugation operator $J$ and 
the modular operator $\Delta$ of Tomita-Takesaki theory.
These results are reproduced here in Propositions \ref {prop:props}, \ref {prop:Jsign},
and \ref {modop:prop:app}.

The theory of the modular operator deals with the symmetry which exists under certain conditions
between a von Neumann algebra and its commutant. In the present context it deals with the symmetry
between two subspaces and their orthogonal complements.
The transition from a real vector space $\mathscr V,s$ to a complex Hilbert space $\mathscr H$
turns out to involve a representation of the split-quaternion group, with its elements linked to the
modular conjugation operator $J$ on ${\mathscr V}$ and the complexification operator $\ihat$ on 
${\mathscr V} \times {\mathscr V}$. The relevant operators on 
$\mathscr V \times \mathscr V$ actually involve a central extension of the split-quaternion group,
with the center constructed from $J$.

Section \ref {sect:pairs} considers pairs of subspaces of a real Hilbert space $\mathscr V$.
The modular operator $\Delta$ and the modular conjugation operator $J$ are introduced.
A simple example considers a line and a plane as a pair of subspaces of
the three-dimensional Euclidean space.

Section \ref {sect:complex} explains how a representation of the group
of split quaternions appears in a natural manner.
By considering $\mathscr V \times \mathscr V$, dimensionality is doubled 
and one can add a complex structure to the Hilbert space. 
The well-known problem of Larmor precession is treated as an example to demonstrate the geometry
of a pair of real subspaces in the complex Hilbert space of wave functions.

A final section discusses the paper and 
offers perspectives for further investigations.

\section{Pairs of Subspaces of $\mathscr V$}
\label{sect:pairs}

\subsection{A pair of orthogonal projections on $\mathscr V$}
\label{sect:orthproj}

In this and the next section, $P$ and $Q$ are a pair of orthogonal projections
in a real Hilbert space ${\mathscr V} ,s$.
Starting from Section \ref {sect:twosub} some restraining assumptions are made.

By definition the projections $P$ and $Q$ satisfy $P^2=P=P^\ast$ and $Q^2=Q=Q^\ast$. Here $\ast$ denotes adjoint operation with respect to the inner product $s$ of the given real Hilbert space ${\mathscr V}$ (fixed throughout this paper).

We consider { the} polar decompositions of $P-Q$ and of $P+Q-\Io$:
\begin{align}
P-Q &= JT ; \label{polar1}\\
P+Q-\Io &= KC . \label{polar2} 
\end{align}
Here, the operators $J$ and $K$
are partial isometries on ${\mathscr V}$.
They are self-adjoint operators because $P-Q$ and $P+Q-\Io$ are self-adjoint.
So $J$ and $T$ commute, and $K$ and $C$ commute. 

The non-negative operators $T$ and $C$ are given by
\[
T=\sqrt{(P-Q)^2}
\quad\mbox{ and }\quad
C=\sqrt{(P+Q-\Io)^2}.
\]
They satisfy
\begin{align*}
T^2
&=(P-Q)^2\cr
&=P+Q-PQ-QP
\end{align*}
and
{
\begin{align*}
C^2
&=(P+Q-\Io)^2\cr
&=\Io+PQ+QP-(P+Q)
\end{align*}
}
so that
\[
T^2+C^2=\Io.
\]

Note that $J^2$ is the orthogonal projection onto the range of $T$
and $K^2$ is the orthogonal projection onto the range of $C$.

\subsection{Identities}

A short calculation shows that
\begin{align}
 T^2&=Q\,(\Io-P)\,Q+(\Io-Q)\,P\,(\Io-Q),\cr
 C^2&=P\,Q\,P+(\Io-Q)\,(\Io-P)\,(\Io-Q).
\end{align}
Note that the operators $P-Q$ and $P+Q-\Io$ anti-commute.
This implies that $PQ-QP=(P-Q)(P+Q-\Io)$ anti-commutes with both
$P-Q$ and $P+Q-\Io$.

%

\begin{proposition}
\label{pairs:prop1}
 The operator $PQ-QP$ commutes with the positive operators $C$ and $T$
 and anti-commutes with the (partial) isometries $J$ and $K$.
\end{proposition}

\beginproof

The operator $PQ-QP$ commutes with $C^2=(P-Q)^2$ and with $T^2=(P+Q-\Io)^2$.
Since C and T are positive operators this implies that it commutes
with $C$ and with $T$ --- see e.g.~Theorem 3.35 in Chapter V of \cite {KT66}.
With similar arguments one shows that $|PQ-QP|$ commutes with the operators $C$, $T$ $P-Q$ and $P+Q-\Io$.
This is used in what follows.

Let
\[
PQ-QP=L\, |PQ-QP|
\]
denote the polar decomposition of $PQ-QP$. Then one calculates
\begin{align*}
 (PQ-QP)J
 &=L\,|PQ-QP|\,T^{-1}(P-Q)\cr
 &=LT^{-1}(P-Q)\,|PQ-QP|\cr
 &=T^{-1}L(P-Q)\,|PQ-QP|\cr
 &=-J(PQ-QP).
\end{align*}
Here, it is used that $L$ commutes with $T^2$ and hence with $T^{-1}$.

The proof that $PQ-QP$ anti-commutes with $K$ is similar.

\endproof

\subsection{Anti-commuting partial isometries}

First a lemma is needed.

\begin{lemma}
\label{lemma1}
Any $\xvec $ belonging to the null space of the operator $T$ or of the operator $C$,
satisfies
\be
\label {ide}
(KJ+JK)\xvec =0.
\ee
\end{lemma}

\beginproof

Consider first the case that $\xvec $ belongs to the null space of the operator $C$, 
i.e.~$C\xvec =0$. Note that this implies that $K\xvec =0$ as well.

Because $T=\sqrt{\Io-C^2}$ this implies that $T\xvec =\xvec $.
It then follows that
\begin{align*}
CKJ\xvec &=KCJT\xvec =(P+Q-\Io)(P-Q)\xvec \cr
&=-(P-Q)(P+Q-\Io)\xvec =-JTCK\xvec =-JCTK\xvec =0.
\end{align*}
This shows that the vector $KJ\xvec $ belongs to the null space of the operator $C$.
From the definition of the polar decomposition $P+Q-1=KC$ it then follows that
the partial isometry $K$ vanishes on $KJ\xvec $. Hence one has
$0=K^2J\xvec $. Because $K^2$ projects onto the range of $C$ this implies
that $J\xvec $ belongs to the null space of $C$. 

By assumption $\xvec $ belongs also to the null space of the operator $C$. 
Hence, $K$ vanishes on both $\xvec $ and $J\xvec $. 
This implies in particular that the identity (\ref{ide})
holds when applied to $\xvec $.

The proof in the case that $\xvec $ belongs to the null space of $T$ is similar.
\endproof

\begin{proposition}
\label{theorem1}
The partial isometries $J$ and $K$ satisfy the anti-commutation relation
\be
KJ+JK=0 .
\label {identityJK}
\ee
\end{proposition}

\beginproof

If $\xvec $ belongs to the kernel of either $T$ or $C$ then $(JK+KJ)\xvec =0$ 
follows from Lemma \ref {lemma1}.
It suffices therefore to prove the relation in the case that $u$ and $v$ exist such that
\[
 \xvec =Tu=Cv.
\]
One can take $u$ in the range of $C$ and $v$ in the range of $T$.
One has 
\begin{align*}
 JK\xvec 
 &=JKCv=J\,(P+Q-\Io)v\cr
 &=J\,(P+Q-\Io)C^{-1}Tu\cr
 &=J\,(P+Q-\Io)TC^{-1}u\cr
 &=J\,(P+Q-\Io)\,(P-Q)\,JC^{-1}u\cr
 &=J\,(QP-PQ)\,\,JC^{-1}u.
\end{align*}
Use Proposition \ref {pairs:prop1} to obtain
\[
 JK\xvec=(PQ-QP)C^{-1}T^{-1}x.
\]
In a similar way one shows that
\[
 KJ\xvec=-(PQ-QP)T^{-1}C^{-1}x.
\]
The two expressions together imply that $(JK+KJ)x=0$.

\endproof

\subsection{Two subspaces}
\label{sect:twosub}

Let $\mathscr K$ and $\mathscr L$ be two closed subspaces of the real Hilbert space 
$({\mathscr V},s)$
with the properties that
\be
{\mathscr K}+{\mathscr L}={{\mathscr V}} , \,\,\,\, \mbox{and} \,\,\,\,
{\mathscr K}\cap{\mathscr L}=\{0\}.
\label{twosub:assum}
\ee 
Then every vector $\xvec$ in ${{\mathscr V}}$ can be decomposed in a unique
way as 
\be
\xvec=\yvec+\zvec
\label{decomp}
\ee
with $\yvec$ in $\mathscr K$ and $\zvec$ in $\mathscr L$.
The subspace $\mathscr K$ is said to be complemented by $\mathscr L$ in $\mathscr V$
and $\mathscr V$ is the direct sum of $\mathscr K$ and $\mathscr L$.
See Definition 4.20 of \cite{RW91}.

In Rieffel and van Daele \cite{RvD77}, 
the stronger requirement is sometimes made that the intersection of any two of
${\mathscr K}$, ${\mathscr L}$, ${\mathscr K}^\perp$, ${\mathscr L}^\perp$
is trivial.  
In the terminology of Halmos \cite{HPR69}, 
the pair ${\mathscr K}$, ${\mathscr L}$ is then said to be in ``generic'' position. 
This assumption is not invoked in what follows unless explicitly stated.

Note that (\ref {decomp}) implies that
\be
{\mathscr K}^\perp\cap{\mathscr L}^\perp=\{0\} 
\label{orthogen}
\ee
Indeed, if $\xvec$ belongs to the intersection then $\xvec$ is orthogonal to both $\yvec$ and $\zvec$
because $\yvec$ belongs to ${\mathscr K}$ and $\zvec$ belongs to ${\mathscr L}$.
Hence it is orthogonal to itself and therefore it must vanish.

In addition, one has also that
\[
{\mathscr K}^\perp +{\mathscr L}^\perp={{\mathscr V}}
\]
because by assumption the intersection of $\mathscr K$ and $\mathscr L$ is trivial.

\subsection{The corresponding projections}

Let $P$ denote the orthogonal projection on $\mathscr K$ and 
$Q$ the orthogonal projection on $\mathscr L$.
By the assumption ${\mathscr K}\cap{\mathscr L}=\{0\}$, 
if $(P-Q)x=0$ then $P\xvec=Q\xvec$ belongs to both ${\mathscr K}$ and ${\mathscr L}$
and therefore $Px$ and $Qx$ must vanish. From the uniqueness of the decomposition (\ref{decomp})
it then follows that $x$ vanishes.
This shows that the operator $P-Q$ is invertible.
Hence, also the operator $T$ is invertible and the partial isometry $J$ from the polar decomposition
$P-Q=JT$ introduced in (\ref{polar1}) is an orthogonal operator 
with $J=J^*$ and $J^2=\Io$.

On the other hand, the operator $P+Q-\Io$ may  still be singular. 
In fact this is always the case if the
two subspaces ${\mathscr K}$ and ${\mathscr L}$ differ in dimension.

For further use the following proposition is needed. It reproduces Proposition 2.2 of \cite{RvD77}.

\begin{proposition}
\label{prop:props}
The following statements hold.
\begin{itemize}
 \item [(a) ] The operators $P+Q$ and { $2\Io-(P+Q)$} are strictly positive;
 \item [(b) ] One has { $T=(P+Q)^{1/2}(2\Io-(P+Q))^{1/2}$;}
 The operator $T$ is strictly positive;
 \item [(c) ] $J$ is a self-adjoint orthogonal operator satisfying $J^2=\Io$.
 \item [(d) ] $T$ commutes with $P$, $Q$ and $J$;
 \item [(e) ] $JP=(\Io-Q)J$ and $JQ=(\Io-P)J$.
\end{itemize}

\end{proposition}

\beginproof
\paragraph*{(a)}
The sum of two non-negative operators is non-negative.
Assume now that $(P+Q)\xvec=0$. This implies that $P\xvec=-Q\xvec$
belongs to the intersection ${\mathscr K}\cap{\mathscr L}$.
By assumption it must vanish, i.e.~$P\xvec=Q\xvec=0$.
The latter implies that $\xvec$ belongs to ${\mathscr K}^\perp\cap{\mathscr L}^\perp$.
This intersection is trivial, see (\ref {orthogen}). Hence, it follows that $\xvec=0$.
This shows that $P+Q$ is strictly positive.

The proof that { $2\Io-(P+Q)$} is strictly positive follows in the same way
with $P$ and $Q$ replaced by $\Io-P$, respectively $\Io-Q$.

\paragraph*{(b)}
A short calculation shows that
\begin{align*}
(P+Q)\,(2\Io-(P+Q))
&=
P+Q-PQ-QP\cr
&=(P-Q)^2\cr
&=T^2.
\end{align*}
Take the square root to obtain $T=(P+Q)^{1/2}(2\Io-(P+Q))^{1/2}$.

Assume now that $T\xvec=0$. This implies $(P-Q)\xvec=0$.
From $P\xvec=Q\xvec$ one deduces that $\xvec=0$ in the same way as in the proof of (a).
Hence, $T$ is strictly positive.

\paragraph*{(c)}
See Section \ref {sect:orthproj}. $J^2=\Io$ follows because the range of $T$ is dense in ${{\mathscr V}}$.

\paragraph*{(d)}
See Section \ref {sect:orthproj}.

\paragraph*{(e)}
One has
\begin{align}
&TJP=JTP=(P-Q)P=P-QP\quad\mbox{ and }\cr
&T(\Io-Q)J=(\Io-Q)TJ=(\Io-Q)(P-Q)=P-QP.
\end{align}
Hence, one has $TJP=T(\Io-Q)J$. Since $T$ is invertible it follows that $JP=(\Io-Q)J$.

The adjoint of $JP=(\Io-Q)J$ is $J(\Io-Q)=PJ$. This shows the second relation.

\endproof

The following result reproduces Proposition 2.3 of \cite{RvD77}.

\begin{proposition}
\label{prop:Jsign}
The operator $J$ is the unique self-adjoint orthogonal operator
on the real Hilbert space ${\mathscr V},s$ with the following properties
\begin{itemize}
 \item [(a) ] $J{\mathscr K}={\mathscr L}^\perp$  and  $J{\mathscr L}={\mathscr K}^\perp$;
 \item [(b) ] If $\yvec\in{\mathscr K}$ and $\zvec\in{\mathscr L}$ then one has
 \[
 s(J\yvec,\yvec)\ge 0, \,\,\,\,\,\,\,\, s(J\zvec,\zvec)\le 0
 \]
 and
 \[
 s(J\yvec,\zvec)=s(J\zvec,\yvec)=0.
 \]
\end{itemize}

\end{proposition}

\beginproof
\paragraph*{(a)} This follows immediately from item (e) of Proposition \ref {prop:props}.

\paragraph*{(b)}
One has
\[
PJP=P(\Io-Q)J=P(P-Q)J=PJTJ=PT=PTP\ge 0.
\]
Here use is made of item (e) of Proposition \ref {prop:props}, of $P^2=P$, of the
fact that $T$ commutes with $P$ and with $J$ --- see item (d) of Proposition \ref {prop:props} ---
and of the fact that $T$ is strictly positive.
From $PJP\ge 0$ one obtains 
\[
s(J\yvec,\yvec)=s(JP\yvec,P\yvec)=s(PJP\yvec,\yvec)\ge 0.
\]

Similarly,
\[
QJQ=Q(\Io-P)J=Q(Q-P)J=-QJTJ=-QT=-QTQ\le 0.
\]
From $QJQ\le 0$ one obtains
\[
s(J\zvec,\zvec)=s(JQ\zvec,Q\zvec)=s(QJQ\zvec,\zvec)\le 0.
\]

Finally, $s(J\yvec,\zvec)=s(J\zvec,\yvec)=0$ follows from item (a) of the proposition.

The above arguments show the existence of an operator $J$ satisfying both (a) and (b).
For the proof of the uniqueness we refer to the proof in \cite{RvD77}.

\endproof

\subsection{The symmetric subspace}
\label{sect:symsub}
Introduce yet another subspace of $\mathscr V$ defined by
\[
{\mathscr S}=\{x:\, Tx=x\}.
\]
Note that ${\mathscr S}$ is the null space of the operator $\Io-T$, hence also of
the operator $(\Io+T)(\Io-T) = \Io-T^2=C^2$. 
The orthogonal complement ${\mathscr S}^\perp$ is the range of the operator $C$.
Hence, $K^*K=K^2$ is the orthogonal projection onto ${\mathscr S}^\perp$. 

\begin{proposition}
\label{symm:prop:props}
 One has 
 \begin{itemize}
  \item [\,(a)\,] $J{\mathscr S}={\mathscr S}$;

  \item [\,(b)\,] If $x$ belongs to $\mathscr S$ and  $x=y+z$ is the unique decomposition of $x$ into an 
  element $y$ of ${\mathscr K}$ and an element $z$ of ${\mathscr L}$, then both $y$ and $z$
  belong to ${\mathscr S}$;
  
  \item [\,(c)\,] If $y$ belongs to ${\mathscr S}\cap{\mathscr K}$ then 
  one has $JQy=-Qy$ and $PQy=0$
  and $Qy$ belongs to ${\mathscr S}$;
  Similarly, if $z$ belongs to ${\mathscr S}\cap{\mathscr L}$ then 
  one has $JPz=Pz$ and $QPz=0$
  and $Pz$ belongs to ${\mathscr S}$;
  
  \item [\,(d)\,] ${\mathscr S}=  ({\mathscr L}^\perp\cap{\mathscr K})
  + ({\mathscr K}^\perp\cap{\mathscr L}$);

    \item [\,(e)\,] ${\mathscr S}\cap{\mathscr K}={\mathscr L}^\perp\cap{\mathscr K}$ and
    ${\mathscr S}\cap{\mathscr L}={\mathscr K}^\perp\cap{\mathscr L}$.
 \end{itemize}
\end{proposition}

\beginproof

\paragraph*{(a)}

Item (d) of Proposition \ref{prop:props} states that $T$ commutes with $J$.
Hence one has
\[
TJx=JTx=Jx
\]
for $x \in \mathscr S$. So $Jx \in \mathscr S$ for any $x\in \mathscr S$. That is, $J {\mathscr S} \subset {\mathscr S} $. 

Now $J^2=\Io$. So $J{\mathscr S}\subset {\mathscr S}$ leads to ${\mathscr S}\subset J{\mathscr S}$.
Therefore $J {\mathscr S} = {\mathscr S} $. 

\paragraph*{(b)}

Item (d) of Proposition \ref{prop:props} states that $T$ commutes with both $P$ and $Q$.
Hence, $Ty$ belongs to $\mathscr K$ and $Tz$ belongs to $\mathscr L$.
From
\[
y+z=x=Tx=Ty+Tz
\]
and the unicity of the decomposition into an element of $\mathscr K$ and an element of $\mathscr L$
it then follows that $Ty=y$ and $Tz=z$, i.e.~both $y$ and $z$ belong to $\mathscr S$.

\paragraph*{(c)}

Take  $y$ in ${\mathscr S}\cap{\mathscr K}$.
Use $J{\mathscr K}={\mathscr L}^\perp$, which is item (a) of Proposition \ref {prop:Jsign},
to obtain
\[
y=J(P-Q)y=J(\Io-Q)y=PJy.
\]
The latter implies
\[
Qy=(\Io-J)y
\]
and 
\[
JQy=-(\Io-J)y=-Qy.
\]
Use this and $JP=(\Io-Q)J$, which item (e) of Proposition \ref {prop:Jsign},
to calculate
\begin{align*}
TQy&=J(P-Q)Qy\cr
&=
JPQy+Qy\cr
&=
(\Io-Q)JQy+Qy\cr
&=
-(\Io-Q)Qy+Qy\cr
&=
Qy.
\end{align*}
This shows that $Qy$ belongs to $\mathscr S$.

Finally, calculate
\begin{align*}
y&=(P-Q)^2y\cr
&=Py+Qy-PQy-QPy\cr
&=(\Io-Q)y+(\Io-P)Qy.
\end{align*}
Compare this with 
\[
y=(\Io-Q)y+Qy
\]
to conclude that $PQy=0$. 

Similarly take $z$ in ${\mathscr S}\cap{\mathscr L}$.
Then one has
\[
z=J(P-Q)z=-J(\Io-P)z.
\]
This implies
\[
Pz=(\Io+J)z
\]
and
\[
JPz=(\Io+J)z=Pz.
\]
Use this to calculate
\begin{align*}
TPz&=J(P-Q)Pz\cr
&=Pz-(\Io-P)JPz\cr
&=Pz.
\end{align*}
This shows that $Pz$ belongs to $\mathscr S$.

The proof that $QPz=0$ is similar to the proof that $PQy=0$.

\paragraph*{(d)}
Take $x$ in ${\mathscr K}^\perp\cap{\mathscr L}$.
It satisfies $Px=0$ and $Qx=x$ so that
\[
Tx=J(P-Q)x=-Jx.
\]
From Proposition \ref{prop:Jsign} it follows that $Jx$ does also belong to 
${\mathscr K}^\perp\cap{\mathscr L}\subset{\mathscr S}$.
Hence one obtains
\[
T^2x=-TJx=-J(-Jx)=x.
\]
This shows that $x$ is an eigenvector of $T^2$ with eigenvalue 1.
Hence it is also an eigenvector of $T$ with eigenvalue 1. This shows that 
$({\mathscr K}^\perp\cap{\mathscr L})\subset{\mathscr S}$.

The proof that $({\mathscr L}^\perp\cap{\mathscr K})\subset{\mathscr S}$ holds 
is similar. One concludes that
\[
({\mathscr K}^\perp\cap{\mathscr L})+({\mathscr L}^\perp\cap{\mathscr K})\subset{\mathscr S}.
\]

To show equality take an arbitrary $x$ in $\mathscr S$ and decompose it as
$x=y+z$ with $y$ in $\mathscr K$ and $z$ in $\mathscr L$.
Let us prove that $y$ belongs also to ${\mathscr L}^\perp$
and $z$ to ${\mathscr K}^\perp$.
By item (b) of the proposition it suffices to prove that $Ty=y$ with $y$ in ${\mathscr K}$
implies that $y$ belongs to ${\mathscr L}^\perp$
and to prove a similar statement when $Tz=z$ with $z$ in ${\mathscr L}$.

By item (c) the vector $Qy$ satisfies $PQy=0$. Hence it belongs to  ${\mathscr K}^\perp\cap{\mathscr L}$.
On the other hand $PQy=0$ implies that
\[
P(\Io-Q)y=y-PQy=y\in{\mathscr K}.
\]
Hence, $(\Io-Q)y$ belongs to ${\mathscr L}^\perp\cap{\mathscr K}$.
The two statements together show that
$y=Qy+(\Io-Q)y$ belongs to 
\[
({\mathscr K}^\perp\cap{\mathscr L}) +({\mathscr L}^\perp\cap{\mathscr K}).
\]
From the unicity of the decomposition of $y$ into an element of $\mathscr K$
and an element of $\mathscr L$ it then follows that $Qy=0$.

\paragraph*{(e)}
The inclusion 
\[
({\mathscr S}\cap{\mathscr K}) \supset ({\mathscr L}^\perp\cap{\mathscr K})
\]
is clear. Take $\xvec$ in ${\mathscr S}\cap{\mathscr K}$. By item (d) it can be written as
$\xvec=\yvec+\zvec$ with $\yvec$ in ${\mathscr L}^\perp\cap{\mathscr K}$ and 
$\zvec$ in ${\mathscr K}^\perp\cap{\mathscr L}$. The decomposition of $\xvec$
as a sum of an element in $\mathscr K$ and an element of $\mathscr L$ is unique.
Because $\xvec$ belongs to $\mathscr K$ this implies that $\xvec=\yvec$ and $\zvec=0$.
This shows the inclusion in the other direction.

The proof of ${\mathscr S}\cap{\mathscr L}={\mathscr K}^\perp\cap{\mathscr L}$
is similar.

\endproof

The proof of the following result is straightforward.

\begin{corollary}
 If the subspaces $\mathscr K = P\mathscr V$ and $\mathscr L = Q \mathscr V$ are in generic position,
 then the subspace
 $\mathscr S$ is trivial: $\mathscr S= \{ 0 \}$.
 On the other hand, if $\mathscr K$ and $\mathscr L$ are ortho-complements:
 $$
 \mathscr K =  (\mathscr L)^\perp, \,\,\,\,  \mathscr L =  (\mathscr K)^\perp, 
 $$
 then one has $P+Q=\Io$, and $\mathscr S$ equals the full space $\mathscr V$: $\mathscr S = \mathscr V$.
\end{corollary}

The following result makes it clear that in many cases one can make the assumption
without loss of generality that the two subspaces $\mathscr K$ and $\mathscr L$
are in generic position w.r.t.~each other.
The above corollary shows that the restriction of $\mathscr K$ and $\mathscr L$ to $\mathscr S$
makes up an orthogonal decomposition of $\mathscr S$.
The proposition that follows shows that the restriction of $\mathscr K$ and $\mathscr L$ to ${\mathscr S}^\perp$
forms two subspaces in generic position w.r.t.~each other.

\begin{proposition}
Take $x$ in the range ${\mathscr S}^\perp$ of the operator $\Io-T$ and let $x=y+z$
be the unique decomposition of $x$ as a sum of an element $y$ in $\mathscr K$ and an element 
$z$ of $\mathscr L$.
Then $y$ and $z$ belong to ${\mathscr S}^\perp$.
\end{proposition}

\beginproof
It suffices to prove that $y$ belongs to ${\mathscr S}^\perp$.

Take any $u$ in ${\mathscr S}$ and decompose it as the sum $u=v+w$
with $v$ in ${\mathscr K}$ and $w$ in ${\mathscr L}$.
One has
\begin{align*}
0&=s(x,v)\cr
&=s(x,Tv)\cr
&=s(Tx,v)\cr
&=s(Ty,v)+s(Tz,v).
\end{align*}
Note that $s(Tz,v)$ vanishes.
Indeed, it equals $s(z,Tv)$ and $Tv=v$ is orthogonal to ${\mathscr L}$.
Hence, one has
\[
0=s(Ty,v)=s(y,Tv)=s(y,v).
\]
By the previous proposition $v$ belongs to ${\mathscr L}^\perp\cap{\mathscr K}$
and $w$ belongs to ${\mathscr K}^\perp\cap{\mathscr L}$. Hence, one has
\[
s(y,u)=s(y,v).
\]
By the previous result one concludes that $y$ is orthogonal to the arbitrary
element $u$ of $\mathscr S$ and hence belongs to its orthogonal complement ${\mathscr S}^\perp$.
\endproof

\subsection{The modular operator}

Item (a) of Proposition \ref {prop:props} states that the operator $P+Q$ is strictly positive.
Hence an eventually unbounded operator $\Delta$ is defined by
\[
\Delta=\frac{2\Io-(P+Q)}{P+Q}
\]
where the right-hand side is 
the function $f(P+Q)$ with $f(u)=(2-u)/u$.

Because also { $2\Io-(P+Q)$} is strictly positive one can conclude that $\Delta$ is a strictly positive operator.
It is shown below that the operator $\Delta$
generalizes the modular operator of Tomita-Takesaki theory \cite{TM70}.

Introduce the map $S$ defined by
\be
S(\yvec+\zvec)=\yvec-\zvec,
\qquad \yvec\in{\mathscr K}, \,\, \zvec\in{\mathscr L}.
\label{modop:Sdef}
\ee
It is well-defined because the decomposition (\ref{decomp}) is unique.
It is linear. To show this use again that the decomposition (\ref{decomp}) is unique.

The following proposition reproduces the results of the Appendix of \cite {RvD77}.

\begin{proposition}
\label{modop:prop:app}
One has
\begin{itemize}
 \item [(a) ] $SP=P$ and $SQ=-Q$;
 \item [(b) ] $S(P+Q)=P-Q$ and $S(P-Q)=P+Q$;
 \item [(c) ] $S^*S=\Delta$;
 \item [(d) ] { $(2\Io-(P+Q))S=P-Q$;}
 \item [(e) ] $S^*J=JS$; 
 \item [(f) ]  The polar decomposition of the operator $S$ is given by $S=J\Delta^{1/2}$.
\end{itemize}
\end{proposition}

\beginproof
\paragraph*{(a)}
This follows from $SPx=Px$ and $SQx=-Qx$, which is valid for all $x$ in ${{\mathscr V}}$.

\paragraph*{(b)}
This follows immediately from (a).

\paragraph*{(c)}

Choose any $\xvec$ and $\xvec'$ and write $\xvec=\yvec+\zvec$ and $\xvec'=\yvec'+\zvec'$
with $\yvec$ and $\yvec'$ in ${\mathscr K}$ and $\zvec$ and $\zvec'$ in $\mathscr L$.
One has
\begin{align*}
s(S(P+Q)x,S\xvec')
&=
s(Px-Qx,\yvec'-\zvec')\cr
&=
s(\yvec+P\zvec-Q\yvec-\zvec,\yvec'-\zvec')\cr
&=
s(\yvec,\yvec')+s(\zvec,\zvec')-s(P\zvec,\zvec')-s(Q\yvec,\yvec')\cr
&=
s((2\Io-(P+Q))x,\xvec').
\end{align*}
This shows that { $S^*S(P+Q)=2\Io-(P+Q)$.}
Divide by $(P+Q)$ to obtain $S^*S=\Delta$.

\paragraph*{(d)}
For any $x$ is
\begin{align*}
(2\Io-(P+Q))Sx
&=
(2\Io-(P+Q))(\yvec-\zvec)\cr
&=
\yvec-\zvec+P\zvec-Q \yvec\cr
&=
(P-Q)x.
\end{align*}

\paragraph*{(e)}
For any $x$ and $\xvec'$ is
\begin{align*}
s(S^*Jx,\xvec')
&=
s(Jx,\yvec'-\zvec')\cr
&=
s(J\yvec,\yvec')-s(J\zvec,\zvec')
\end{align*}
because
\[
s(J\zvec,\yvec')=s(J\yvec,\zvec')=0.
\]
The latter is a consequence of item (a) of Proposition \ref {prop:Jsign}.

On the other hand one has
\begin{align*}
s(JSx,\xvec')
&=
s(J\yvec,\xvec')-s(J\zvec,\xvec')\cr
&=
s(J\yvec,\yvec')-s(J\zvec,\zvec')
\end{align*}
with the same argument as in the previous calculation.
The two calculations together prove that $S^*J=JS$.

\paragraph*{(f)}

From the definition of the operator $T$ and (d) of the present proposition one obtains
\begin{align*}
T
&=J(P-Q)\cr
&=J(2\Io-(P+Q))S.
\end{align*}
Next use (e) of Proposition \ref {prop:props} to obtain
\[
T=(P+Q)JS.
\]
Multiply with $(P+Q)^{-1/2}$ and use (b) of Proposition  \ref {prop:props}
to obtain
\begin{align*}
(P+Q)^{1/2}JS
&=(2\Io-(P+Q))^{1/2}\cr
&=(P+Q)^{1/2}\Delta^{1/2}.
\end{align*}
Because $P+Q$ is strictly positive and $J^2=\Io$ one obtains
$S=J\Delta^{1/2}$.
The operator $J$ is an isometry and $\Delta^{1/2}$ is non-negative.
Hence , $S=J\Delta^{1/2}$ is the polar decomposition of $S$.
\endproof

From $S^2=1$ it follows that
\[
\Delta^{-1/2}=J\Delta^{1/2}J.
\]
The adjoint $S^*$ is then given by
\[
S^*=\Delta^{1/2}J=J\Delta^{-1/2}.
\]

\subsection{Example: Euclidean space}
\label{sect:eg:eucl}

Let ${\mathscr V}=\Ro^3$ and choose ${\mathscr K}$ and ${\mathscr L}$ to be
\[
{\mathscr K}=\{x \in \Ro^3:\, x_1=x_2=x_3\}
\quad\mbox{ and }\quad
{\mathscr L}=\{x \in \Ro^3:\, x_1=0\}.
\]
So ${\mathscr K}$ is a line given by  $x_1=x_2=x_3$, whereas ${\mathscr L}$ is a plane given by $x_1=0$.
The intersection of the two subspaces is trivial, and the sum
${\mathscr K}+{\mathscr L}$ spans all of ${\mathscr V}$.
The inner product $s$ is the usual
\[
s(x,y)=\sum_{i=1}^3x_iy_i.
\]
The plane ${\mathscr L}$ is not orthogonal to the line ${\mathscr K}$.
The orthogonal complements are given by
\begin{align*}
{\mathscr K}^\perp&=\{x \in \Ro^3:\, x_1+x_2 +x_3=0\},\cr
{\mathscr L}^\perp&=\{x \in \Ro^3:\, x_2=x_3=0\}.
\end{align*}
The intersection ${\mathscr K}\cap{\mathscr L}^\perp$ evaluates to $\{0\}$.
The intersection
\[
{\mathscr L}\cap{\mathscr K}^\perp=\{x \in \Ro^3:\,x_1=x_2+x_3=0\}
\]
is one-dimensional. So the subspace $\mathscr S$ satisfies
$$ {\mathscr S} = {\mathscr L}\cap{\mathscr K}^\perp +
{\mathscr K}\cap{\mathscr L}^\perp = {\mathscr L}\cap{\mathscr K}^\perp + \{0 \}
= {\mathscr L}\cap{\mathscr K}^\perp.$$ 
This gives
$$
{\mathscr S} = \{x \in \Ro^3:\,x_1=x_2+x_3=0\}
.$$
For this example, the subspaces ${\mathscr K}$ and ${\mathscr L}$ are {\it not} in generic position.

The matrix $P$ projecting on $\mathscr K$ is explicitly given by
\[
P=
\frac 13
\left(\begin{array}{lcr}
 1 &1 &1\cr
 1 &1 &1\cr
 1 &1 &1\cr 
\end{array}\right).
\]
The matrix $Q$ projecting on $\mathscr L$ is given by
\[
Q=
\left(\begin{array}{lcr}
 0 &0 &0\cr
 0 &1 &0\cr
 0 &0 &1\cr 
\end{array}\right) .
\]
Hence, $P-Q$ is given by
\[
P-Q=
\frac 13
\left(\begin{array}{lcr}
 1 &1 &1\cr
 1 &-2 &1\cr
 1 &1 &-2\cr 
\end{array}\right).
\]

The operator $T$ is determined by $T^2=(P-Q)^2$.
One has
\begin{align*}
T^2&=P+Q-PQ-QP\cr
&=
\frac 13
\left(\begin{array}{lcr}
 1 &0 &0\cr
 0 &2 &-1\cr
 0 &-1 &2\cr 
\end{array}\right)
\end{align*}
The eigenvalues of $T^2$ are $1/3$, $1$ and $1/3$
with corresponding eigenvectors
\[
\evec_1=
\left(\begin{array}{c}
1\\0\\0
\end{array}\right),
\quad
\evec_2=\frac{1}{\sqrt 2}
\left(\begin{array}{c}
0\\1\\-1
\end{array}\right),
\quad
\evec_3=\frac{1}{\sqrt 2}
\left(\begin{array}{c}
0\\1\\1
\end{array}\right).
\]
Calculating the square root $T$ of $T^2$, one finds
\[
T=
\frac 1{2\sqrt{3}}
\left(\begin{array}{lcr}
 2 &0 &0\cr
 0 &1+\sqrt 3 &1-\sqrt 3\cr
 0 &1-\sqrt 3 &1+\sqrt 3\cr 
\end{array}\right).
\]
The inverse $T^{-1}$ of $T$ is given by
\begin{align*}
T^{-1}
&=\sqrt 3\,\evec_1 + \evec_1+\sqrt 3\, \evec_1\cr
&=
\frac 12
\left(\begin{array}{lcr}
 2\sqrt 3 &0 &0\cr
 0 &1+\sqrt 3 &-1+\sqrt 3\cr
 0 &-1+\sqrt 3 &1+\sqrt 3\cr 
\end{array}\right).
\end{align*}
The inverse is used to calculate the isometry $J$ in the polar decomposition $P-Q=JT$.
One finds
\begin{align*}
J&=(P-Q)T^{-1}\cr
&=
\frac 16
\left(\begin{array}{lcr}
 2\sqrt 3 &2\sqrt 3 &2\sqrt 3\cr
 2\sqrt 3  &-3-\sqrt 3 &3-\sqrt 3\cr
 2\sqrt 3  &3-\sqrt 3 &-3-\sqrt 3\cr 
\end{array}\right).
\end{align*}
It satisfies $J=J^*$ and $J^2=\Io$, as it should be.
One verifies that $TJ=P-Q=JT$ holds.

The matrix $P+Q-\Io$ is given by
\[
P+Q-\Io
=
\frac 13
\left(\begin{array}{lcr}
 -2 &1 &1\cr
 1 &1 &1\cr
 1 &1 &1\cr 
\end{array}\right).
\]
Its square equals
\[
(P+Q-\Io)^2
=
\frac 13
\left(\begin{array}{lcr}
 2 &0 &0\cr
 0 &1 &1\cr
 0 &1 &1\cr 
\end{array}\right).
\]
The square root $C$ of $C^2 \equiv (P+Q-\Io)^2$ can then be calculated. One finds
\[
C=
\frac 1{\sqrt{6}}
\left(\begin{array}{lcr}
 2 &0 &0\cr
 0 &1 &1\cr
 0 &1 &1\cr 
\end{array}\right).
\]
The polar decomposition of $P+Q-\Io$ is given by
\[
P+Q-\Io=KC
\]
with $K$ the partial isometry determined by
\begin{align*}
K\evec_1&=
\frac{1}{\sqrt 6}
\left(\begin{array}{c}
      -2\\1\\1
\end{array}\right),\cr
K\evec_2&=0,\cr
K\evec_3&=
\frac{1}{\sqrt 3}
\left(\begin{array}{c}
      1\\1\\1
\end{array}\right).
\end{align*}
The solution is
\[
K=
\frac{1}{\sqrt 6}
\left(\begin{array}{lcr}
 -2 &1 &1\cr
 1 &1 &1\cr
 1 &1 &1\cr 
\end{array}\right).
\]
Its square
\[
K^2=\frac{1}{2}
\left(\begin{array}{lcr}
 2 &0 &0\cr
 0 &1 &1\cr
 0 &1 &1\cr 
\end{array}\right).
\]
is the orthogonal projection onto the space ${\mathscr S}^\perp$
spanned by the basis vectors $\evec_1$ and $\evec_3$.

The product $JK$ is  anti-Hermitian. 
One finds
\[
JK=\frac 1{\sqrt 2}
\left(\begin{array}{lcr}
 0 &1 &1\cr
 -1 &0 &0\cr
 -1 &0 &0\cr 
\end{array}\right).
\]
The operator $S$ follows from item (b) of Proposition \ref{prop:props}.
One finds
\begin{align*}
S&=
(P-Q)(P+Q)^{-1}\cr
&=
\frac 13
\left(\begin{array}{lcr}
 1 &1 &1\cr
 1 &-2 &1\cr
 1 &1 &-2\cr 
\end{array}\right)
\left(\begin{array}{lcr}
 5 &-1 &-1\cr
 -1 &1 &0\cr
 -1 &0 &1\cr 
\end{array}\right)\cr
&=
\left(\begin{array}{lcr}
 1 & 0 &0\cr
 2 &-1 &0\cr
 2 &0 &-1\cr 
\end{array}\right).
\end{align*}
The modular operator is given by
\begin{align*}
\Delta
&=S^*S\cr
&=
\left(\begin{array}{lcr}
 1 & 2 &2\cr
 0 &-1 &0\cr
 0 &0 &-1\cr 
\end{array}\right)\,
\left(\begin{array}{lcr}
 1 & 0 &0\cr
 2 &-1 &0\cr
 2 &0 &-1\cr 
\end{array}\right)\cr
&=
\left(\begin{array}{lcr}
 9 & -2 &-2\cr
 -2 &1 &0\cr
 -2 &0 &1\cr 
\end{array}\right).
\end{align*}

\section{Complex Structure on 
{ $\mathscr V \times \mathscr V $}}
\label{sect:complex}

\subsection{Split quaternions}

The operator $J$ describes the symmetry between the pair of subspaces ${\mathscr K}$, ${\mathscr L}$
and the pair ${\mathscr L}^\perp$, ${\mathscr K}^\perp$. 
It is convenient to describe the pairs simultaneously in the product space 
{
${\mathscr V}\times {\mathscr V}$.}
The elements of the latter are denoted $[x,y]$, with $x$ and $y$ in ${\mathscr V}$,
rather than $(x\,\,y)^\tinyT$ or $x\oplus y$ or $x+ iy$.

Introduce the linear map $\hat j$ defined by
\[
\hat j:\,[x,y]\mapsto [Jy,Jx],
\qquad x,y\in {{\mathscr V}}.
\]
It clearly satisfies $\hat j^2=1$ and maps ${\mathscr K}\oplus{\mathscr L}$ onto 
${\mathscr K}^\perp\oplus{\mathscr L}^\perp$.

In addition the space ${{\mathscr V}}$ 
itself has symmetries as well.
In particular, the inversion $x\mapsto -x$ leaves the subspaces invariant.
This is a gauge symmetry because it describes a change of coordinate system rather than
a change of the subspaces. 

Introduce the linear map $\hat k$ defined by
\[
\hat k:\,[x,y]\mapsto [x,-y],
\qquad x,y\in {{\mathscr V}}.
\]
It satisfies $\hat k^2=1$ and $\hat j\hat k+\hat k\hat j=0$. Indeed, one has
\[
\hat j\hat k[x,y]=\hat j[x,-y]=[-Jy,Jx]=-\hat k[Jy,Jx]=-\hat k\hat j[x,y].
\]

Let $\hat i=\hat k\hat j=-\hat j\hat k$. The linear map $\hat i$ satisfies $\hat i^2=-1$ and 
$\hat i\hat j+\hat j\hat i=\hat i\hat k+\hat k\hat i=0$ and
\[
\hat i[x,y]=[Jy,-Jx],
\qquad x,y\in {{\mathscr V}}.
\]

Along with the identity map
$\hat I$ defined by
$ \hat I [x,y] = [x,y] ,$
the eight elements $\pm \hat I,\pm \hat i,\pm \hat j,\pm \hat k$ together form a representation of the dihedral group $D_4$, which can be identified with the so-called split-quaternion.
Note that the symmetry of the square is the natural symmetry for the four objects 
${\mathscr K}$, ${\mathscr L}$, ${\mathscr L}^\perp$, ${\mathscr K}^\perp$.

\subsection{Complexification}

The linear map $\hat i$ defines a complexification of the real Hilbert space $({\mathscr V},s)$.
The resulting complex Hilbert space is denoted $\mathscr H$. For clarity of presentation
it is identified with the product space 
{ ${\mathscr V}\times {\mathscr V}$.}
The complex-valued inner product $(\cdot,\cdot)_s$ of $\mathscr H$ is determined by the requirement that
\[
([x,0],[y,0])_s =s(x,y).
\]
It is an extension of $s$ into a sesquilinear form $\tilde{s}$:
$$
([x,y], [u,v])_s = \tilde{s}(x - iJy, u+i Jv) .
$$

Explicitly written out, using the fact $s(Jy, Jv) = s(y,v)$ due to $J$ being an isometry,
\[
([x,y],[u,v])_s
=s(x,u)+s(y,v)+is(x,Jv)- is(Jy,u).
\]
One verifies that
\begin{align*}
(\hat i[x,y],[u,v])_s
&=
([Jy,-Jx],[u,v])\cr
&=
s(Jy,u)+s(-Jx,v)+is(Jy,Jv)-is(-J^2 x,u)\cr
&=
s(Jy, u) - s(Jx,v) + is(Jy,Jv)  + i s(x,u)  \cr
&= is(x,u) + is(y,v)  -s(x,Jv) + s(Jy,u) \cr
&=
i([x,y],[u,v])_s.
\end{align*}
So $\hat i$ indeed behaves as a real-linear complexification operator
with respect to the sesquilinear inner-product $(\cdot, \cdot)_s$ on $\mathscr H$.

The following result is a direct consequence of Proposition \ref{prop:Jsign}.

\begin{proposition}
If $x$ and $y$ belong to $\mathscr K$ and $u$ and $v$ belong to $\mathscr L$
then the inner product $([x,y],[u,v])_s$ turns out to be real-valued and equals
\[
([x,y],[u,v])_s=s(x,u)+s(y,v).
\]
The inner product $([x,u],[y,v])_s$ turns out to be real as well and is given by
\[
([x,u],[y,v])_s=s(x,y)+s(u,v).
\]
\end{proposition}

The imaginary part of the complex inner product defines a symplectic form $\omega_s$.
It is given by
\[
\omega_s([x,y],[u,v])=s(x,Jv)-s(Jy,u).
\]
The proposition shows that if $x$ and $y$ belong to $\mathscr K$ and $u$ and $v$ belong to $\mathscr L$
then $\omega_s([x,u],[y,v])$ vanishes. In the terminology of \cite{dG06}
this means that the subspaces $\mathscr K$ and $\mathscr L$ form a pair of
Lagrangian planes.

\subsection{Complex-linear operators}

Given two linear operators $A$ and $B$ on ${\mathscr V}$ one can construct a linear operator $[A\sep B]$ on $\mathscr H$
defined by
\[
[A\sep B]=
\left(\begin{array}{lr}
       A &BJ\\
       -JB &JAJ
      \end{array}\right).
\]
One verifies that 
\begin{align*}
[A\sep B]\,\ihat\, [x, y]
&=
\left(\begin{array}{lr}
       A &BJ\\
       -JB &JAJ
      \end{array}\right)
\,\left(\begin{array}{c}
         Jy\\-Jx
        \end{array}\right)\cr
&=
[AJy-Bx, -JBJy-JAx]\cr
&=
\ihat [A\sep B]\, [x, y].
\end{align*}
Hence, the operator $[A\sep B]$ is complex-linear.
Next verify that
\begin{align*}
([A\sep B]\,[x,y],[u,v])_s
&=
([Ax+BJx,-JBx+JAJy],[u,v])_s\cr
&=
s(Ax+BJy,u)+s(-JBx+JAJy,v)\cr
&\quad + is(Ax+BJy,Jv)- is(-Bx+AJy,u)\cr
&=
s(x,A^*u-B^*Jv)+s(y,JA^*Jv+JB^*u)\cr
&\quad
+i s(x,A^*Jv+B^*u)- is(y,-JB^*Jv+JA^*u)\cr
&=
([x,y],[A^*u-B^*Jv, JA^*Jv+JB^*u)_s\cr
&=
([x,y],[A^*\sep -B^*]\,[u,v])_s.
\end{align*}
This shows that $[A^*\sep -B^*]$ is the adjoint  of $[A\sep B]$:
$$
[A\sep B]^\star = [A^*\sep -B^*] .
$$

Note that $\ast$ is the adjoint transformation of operators on ${\mathscr V}$ with respect to $s$, while $\star$ denotes adjoint transformation of operators on ${\mathscr H}$ with respect to $(\cdot, \cdot)_s$.

Note that $[\Io\sep 0]$ is the identity operator $\hat\Io$ on $\mathscr H$.
On the other hand, the square root of $-[\Io\sep 0]$ is $\pm \ihat$
with
\[
\ihat=\left(\begin{array}{lr}
       0 &J\\
       -J &0
      \end{array}\right)
=[0\sep \Io].
\]

The product of two complex-linear operators $[A\sep B]$ and $[C\sep D]$ is given by
\[
[A\sep B]\,[C\sep D]=[AC-BD,AD+BC].
\]

Positive operators are of the form
\begin{align}
[A\sep B]^\star \,[A\sep B]
&=
\left(\begin{array}{lr}
       A^* &-B^*J\\
       JB^* &JA^*J
      \end{array}\right)\,
\left(\begin{array}{lr}
       A &BJ\\
       -JB &JAJ
      \end{array}\right)\cr
&=
[A^*A+B^*B\sep A^*B-B^*A].
\label{posop}
\end{align}

\subsection{ Antilinear operators}
\label{complex_antilinear}

Introduce the notation
\[
[A\antisep B]=
\left(\begin{array}{lr}
       A &BJ\\
       JB &-JAJ
      \end{array}\right).
\]
One verifies that 
\begin{align*}
[A\antisep B]\ihat [x, y]
&=
\left(\begin{array}{lr}
       A &BJ\\
       JB &-JAJ
      \end{array}\right)
\,\left(\begin{array}{c}
         Jy\\-Jx
        \end{array}\right)\cr
&=
[AJy-Bx, JBJy+JAx]\cr
&=
-\ihat [A\antisep B]\, [x, y].
\end{align*}
Hence, the operator $[A\antisep B]$ is { antilinear}.
The adjoint of $[A\antisep B]$ follows from
\begin{align*}
([A\antisep B]\,[x,y],[u,v])_s
&=
([AJy-BJx,-JBy-JAx],[u,v])_s \cr
&=
s(Ax+BJy,u)+s(JBx-JAJy,v)\cr
&+is(Ax+BJy,Jv)-is(Bx-AJy,u)\cr
&=
s(x,A^*u+B^*Jv)+s(y,-JA^*Jv+JB^*u)\cr
& 
+is(x,A^*Jv-B^*u)-is(y,JB^*Jv-JA^*u)\cr
&=
([x,y],[A^*u+B^*Jv,- JA^*Jv+JB^*u)_s \cr
&=
([A^*\antisep B^*]\,[u,v], [x,y])_s.
\end{align*}
Hence, one has 
$$[A\antisep B]^\star=[A^*\antisep B^*]. $$

The operators $\jhat$ and $\khat$ anti-commute with $\ihat$. Hence they are con\-jugate-lin\-ear.
The explicit expressions are
\[
\jhat=\left(\begin{array}{lr}
       0 &J\\
       J &0
      \end{array}\right)
=[0\antisep \Io]
\quad\mbox{ and }\quad
\khat=
\left(\begin{array}{lr}
       \Io &0\\
       0 &-\Io
      \end{array}\right)
=[0\antisep J].
\]

\subsection{Central extension}

Note that all operators of the form $[A\sep 0]$ commute with $\ihat$, $\jhat$ and $\khat$.
In particular, the operator
\[
[J\sep 0]=\left(\begin{array}{lr}
       J &0\\
       0 &J
      \end{array}\right)
\]
is a self-adjoint operator satisfying $[J\sep 0]^2=[\Io\sep 0]$ and commuting with $\ihat$, $\jhat$ and $\khat$.
It defines
a central extension of the split-quaternion representation. It is denoted $\chat$ in what follows,
i.e.
\[
 \hat c=[J\sep 0].
\]

Consider also the operators
\[
\khat:=
[\Io\antisep 0]=
\left(\begin{array}{lr}
       \Io &0\\
       0 &-\Io
      \end{array}\right)
\quad\mbox{ and }\quad
\mhat:=
[J\antisep 0]=
\left(\begin{array}{lr}
       J &0\\
       0 &-J
      \end{array}\right)
\]
and
\[
 \dhat:= [0\sep J]
 =\left(\begin{array}{lr}
       0&\Io\\
       -\Io &0
      \end{array}\right)
\quad\mbox{ and }\quad
\hhat:=
[0\antisep J]
=\left(\begin{array}{lr}
       0&\Io\\
       \Io &0
      \end{array}\right).
\]

The 8 self-adjoint operators
\[
\pm[\Io\sep 0],\,\, \pm[0\sep \Io],\,\, \pm[J\sep 0],\,\, \pm[0\sep J]
\]
and the 8 anti-adjoint operators
\[
\pm[\Io\antisep 0], \,\, \pm[0\antisep \Io], \,\, \pm[J\antisep 0], \,\, \pm[0\antisep J]
\]
together form a group of order 16. It is denoted $\mathbb G$ hereafter.
The reduced group table is given by Table \ref{table:centex:group}.

\begin{table}
\begin{center}
\begin{tabular}{l|cccccccc}
                  &$\hat\Io$             &  $\ihat$                 & $\chat$            &$\dhat$            
                      &$\khat$              &  $\jhat$                  &$\mhat$   & $\hhat$ \\
\hline
$\hat\Io$   &$\hat\Io$        &  $\ihat$         & $\chat$        &$\dhat$
                      &$\khat$  &  $\jhat$    & $\mhat$ &$\hhat$ \\
$\ihat$      &$\ihat$         &$-\hat\Io$         &$\dhat$        &$-\chat$
                      &$-\jhat$  &$\khat$       &$-\hhat$ &$\mhat$  \\
$\chat$     &$\chat$           &$\dhat$             &$\hat\Io$     &$\ihat$
                      &$\mhat$     &$\hhat$        &$\khat$    &$\jhat$\\
$\dhat$     &$\dhat$           &$-\chat$             &$\ihat$        &$-\hat\Io$
                      &$-\hhat$    &$\mhat$        &$-\jhat$&$\khat$\\
$\khat$     &$\khat$  &$\jhat$     &$\mhat$   &$\hhat$
                      &$\hat\Io$        &$\ihat$          &$\chat$        &$\dhat$\\
$\jhat$      &$\jhat$  &$-\khat$   &$\hhat$   &$-\mhat$
                      &$-\ihat$      &$\hat\Io$          &$-\dhat$       &$\chat$\\
$\mhat$    &$\mhat$     &$\hhat$       &$\khat$ &$\jhat$
                     &$\chat$           &$\dhat$             &$\hat\Io$      &$\ihat$\\
$\hhat$     &$\hhat$     &$-\mhat$      &$\jhat$&$-\khat$
                     &$-\dhat$          &$\chat$             &$-\ihat$     &$\hat\Io$\\
\end{tabular}
\caption{Reduced group table}
\label{table:centex:group}
\end{center}
\end{table}

The group $\mathbb G$ is 2-graded. The even elements are $\pm\hat \Io,\, \pm\ihat, \, \pm\chat, \, \pm\dhat$,
the odd elements are $\pm\hhat, \, \pm\jhat, \, \pm\mhat, \, \pm\khat$.
The group of the split-quaternions 
\[
\mathbb N=\{\pm\hat \Io, \, \pm\ihat, \, \pm\jhat, \, \pm\khat\}
\]
is a normal subgroup of $\mathbb G$.
One verifies that the group $\mathbb G$ is the semi-direct product of the
subgroup $\{\hat\Io,\chat\}$ with the normal subgroup $\mathbb N$.
Hence, the group $\mathbb G$ is the central extension of $\mathbb N$ by the central element $\chat$ of $\mathbb G$.

In this group $\mathbb G$, the elements $\pm \ihat$ and $\pm \dhat$ are of order 4, and $(\hat\Io, \ihat, -\hat\Io, -\ihat)$ and $(\hat\Io, \dhat, -\hat\Io, -\dhat)$
are cyclic subgroup ${\mathbb Z}_4$. All other non-identity elements are of order 2. Furthermore, there are 4 sets of split-quaternions
$$
 \{ \hat\Io, \ihat, \jhat, \khat \}, \,\,\,\,
\{ \hat\Io, \ihat, \mhat, \hhat \},  \,\,\,\,
\{ \hat\Io, \dhat, \jhat, \mhat \}, \,\,\,\, 
\{ \hat\Io, \dhat, \khat, \hhat \}.
$$

Table \ref {table:centex} summarizes the notations being used.

\begin{table}

\begin{center}
\begin{tabular}{l|c|c}
\hline
$\hat\Io$      &$[\Io \sep 0]$
    &$\left(\begin{array}{lr}
       \Io &0\\
       0 &\Io
      \end{array}\right)$\\
$\chat$         &$[J\sep 0]$
    &$\left(\begin{array}{lr}
       J &0\\
       0 &J
      \end{array}\right)$\\
$\ihat$         &$[0\sep \Io]$
    &$\left(\begin{array}{lr}
       0 &J\\
       -J &0
      \end{array}\right)$\\
$\jhat$         &$[0\antisep\Io]$
    &$\left(\begin{array}{lr}
       0 &J\\
       J &0
      \end{array}\right)$\\
$\khat$         &$[\Io \antisep 0]$
    &$\left(\begin{array}{lr}
       \Io &0\\
       0 &-\Io
      \end{array}\right)$\\
$\dhat$         &$[0\sep J]$
    &$\left(\begin{array}{lr}
       0 &\Io\\
       -\Io &0
      \end{array}\right)$\\
$\mhat$         &$[J\antisep 0]$
    &$\left(\begin{array}{lr}
       J &0\\
       0 &-J
      \end{array}\right)$\\
$\hhat$         &$[0\antisep J]$
    &$\left(\begin{array}{lr}
       0 &\Io\\
       \Io &0
      \end{array}\right)$\\
\end{tabular}\\
\caption{Notations}
\label{table:centex}
\end{center}
\end{table}

\subsection{Orthogonal projections in { ${\mathscr V} \times {\mathscr V}$}}
\label{sect:complsub}

Identify $\mathscr V \times \mathscr V$ with the direct sum $\mathscr V \oplus \mathscr V$
 of the real Hilbert space $\mathscr V,s$ with itself.
 Then one can consider the following subspaces:
\[
\Kplus={\mathscr K}\oplus{\mathscr K}^\perp
\quad\mbox{ and }\quad
\Lplus={\mathscr L}\oplus{\mathscr L}^\perp.
\]
These subspaces are the range of a pair of orthogonal projections $\phat$ and $\qhat$ on 
{ $\mathscr V \times \mathscr V$}:
\[
\phat=
\left(\begin{array}{lr}
       P &0\\
       0 &\Io-P
      \end{array}\right)
      \quad\mbox{ and }\quad
\qhat=
\left(\begin{array}{lr}
       Q &0\\
       0 &\Io-Q
      \end{array}\right) ,
\]
where $P, Q$ are the pair of projections $P, Q$ on $\mathscr V$. The $\phat, \qhat$ projections satisfy $\phat^2 = \phat = \phat^\star$ and $\qhat^2 = \qhat = \qhat^\star$.

The ortho-complement of $\Kplus$ and $\Lplus$ with respect to the real-valued
inner-product
of $\mathscr V \oplus \mathscr V$
are denoted $(\Kplus)^\perp$ and $(\Lplus)^\perp$. They correspond to the following pair of projectors:
\[
\hat{\Io} - \phat=
\left(\begin{array}{lr}
       \Io - P &0\\
       0 & P
      \end{array}\right)
      \quad\mbox{ and }\quad
\hat{\Io}- \qhat=
\left(\begin{array}{lr}
       \Io - Q &0\\
       0 & Q
      \end{array}\right) 
\]
where
\[
\hat{\Io} =
\left(\begin{array}{lr}
       \Io  &0\\
       0 & \Io
      \end{array}\right) 
\]
is the identity operator on { $\mathscr V \times \mathscr V$}.

The following two propositions show that the spaces $\Kplus$, $\Lplus$, $(\Kplus)^\perp$, $(\Lplus)^\perp$, as subspaces of {  $\mathscr V \times \mathscr V$}, 
are mutually in general position. The spaces $\Kplus$ and $\Lplus$ are in generic position
in Halmos' terminology.

\begin{proposition}
The spaces $\Kplus$ and $\Lplus$ satisfy
\[
\Kplus +\Lplus=\Kplus^\perp +\Lplus^\perp={\mathscr V}\times {\mathscr V}.
\]
The intersection of any two of the subspaces $\Kplus$, $\Lplus$, $(\Kplus)^\perp$, $(\Lplus)^\perp$
is trivial.
\end{proposition}

\beginproof

By assumption is ${\mathscr V}={\mathscr K}+{\mathscr L}$
and $\{0\}={\mathscr K}\cap{\mathscr L}$, which implies ${\mathscr V}={\mathscr K}^\perp+{\mathscr L}^\perp$.
Hence, an arbitrary element $[\xvec,\yvec]$ of ${\mathscr V}\times {\mathscr V}$
can be written as
\[
 [\xvec,\yvec]=[\uvec+\vvec,\zvec+\wvec]=[\uvec,\zvec]+[\vvec,\wvec]
\]
with $\uvec$ in $\mathscr K$ and $\vvec$ in $\mathscr L$, $\zvec$ in ${\mathscr K}^\perp$
and $\wvec$ in ${\mathscr L}^\perp$.
Note that $[\uvec,\zvec]$ belongs to $\Kplus$ and $[\vvec,\wvec]$ belongs to $\Lplus$.
This shows that $\Kplus +\Lplus={\mathscr V}\times {\mathscr V}$.
As a consequence one has also $\Kplus^\perp\cap\Lplus^\perp=\{0\}$.

Next assume that $[x,y]$ belongs to $\Kplus\cap\Lplus$.
Then $x$ belongs to ${\mathscr K}\cap{\mathscr L}$
and $y$ to ${\mathscr K}^\perp\cap{\mathscr L}^\perp$.
By assumption ${\mathscr K}$ and ${\mathscr L}$ are in general position.
Hence, it follows that $x=0$. It is noted in Section \ref {sect:twosub}
that the pair ${\mathscr K}^\perp$, ${\mathscr L}^\perp$
is also in general position. This implies that $y=0$ as well.
One concludes that $\Kplus\cap\Lplus$ is trivial.
This implies also that $\Kplus^\perp +\Lplus^\perp={\mathscr V}\times {\mathscr V}$.

Next assume that $[x,y]$ belongs to $\Kplus\cap\Lplus^\perp$.
Then one has 
\[
([x,y],[u,v])_s=0
\]
for any $[u,v]$ in $\Lplus$.
From the definition of the complex inner product in 
{ ${\mathscr V}\times {\mathscr V}$}
one deduces that 
\[
s(x,Jv)=s(Jy,u).
\]
Take $u=aJy$ and $v=bJx$ with $a$ and $b$ arbitrary real numbers.
Note that $x$ belongs to $\mathscr K$ and $y$ belongs to ${\mathscr K}^\perp$.
By Proposition \ref{prop:Jsign}$v$ belongs to ${\mathscr L}^\perp$ and $u$ belongs to ${\mathscr L}$
so that $[u,v]$ belongs to $\Lplus$. Therefore one has 
\[
bs(x,x)=as(y,y).
\]
Since $a$ and $b$ are arbitrary this implies that
\[
s(x,x)=s(y,y)=0.
\]
This shows that the intersection $\Kplus\cap\Lplus^\perp$ is trivial.

The proof that $\Kplus^\perp\cap\Lplus=\{0\}$ is similar.

\endproof

\subsection{Polar decomposition}

The following result shows that the operator $\phat-\qhat$
is the product of the conjugate-linear operator $\mhat$ with the complex-linear operator $[T\sep 0]$.
 
\begin{proposition}
The polar decomposition of the operator $\phat-\qhat$ is given by
\[
\phat-\qhat=[J \antisep 0] \,[T\sep 0].  
\]
where $[T\sep 0]$ and $[J \antisep 0]$ are given by 
\[
[T\sep 0] =
\left(\begin{array}{lr}
       T &0\\
       0 &T
      \end{array}\right) , \,\,\,\,\,\,\,\,
  [J\antisep 0] =
\left(\begin{array}{lr}
       J &0\\
       0 & -J
      \end{array}\right) = \mhat .    
\]
 
\end{proposition}

\beginproof
First, item (d) of Proposition \ref{prop:props} shows that the operator $T$ commutes with $J$.
Use this to obtain
\begin{align*}
[T\sep 0]\,[x,y]
&=
\left(\begin{array}{lr}
       T &0\\
       0 &JTJ
      \end{array}\right)\,
\left(\begin{array}{c}
         x\\y
\end{array}\right)\cr
&=
[Tx,JTJy]\cr
&=
[Tx,Ty].
\end{align*}
Because $T$ is a positive operator the condition (\ref {posop}) is satisfied
so that also $[T\sep 0]$ is a positive operator.
The operator $[T\sep 0]$ shares these properties with the operator $[C\sep 0]$.

For any $[x,y]$ and $[u,v]$ in 
{ ${\mathscr V}\times {\mathscr V}$} one has
\begin{align*}
&
((\phat-\qhat)[x,y],(\phat-\qhat)[u,v])_s \cr
&=
([Px,(\Io-P)y]-[Qx,(\Io-Q)y],
[Pu,(\Io-P)v]-[Qu,(\Io-Q)v])_s \cr
&=
([(P-Q)x,-(P-Q)y][(P-Q)u,-(P-Q)v])_s \cr
&=
s(((P-Q)x,(P-Q)u)+s((P-Q)y,(P-Q)v)\cr
&\quad
-is((P-Q)x,J(P-Q)v)+is(J(P-Q)y,(P-Q)u)\cr
&=
s(Tx,Tu)+s(Ty,Tv)
-is(JTx,Tv)+is(Ty,JTu)\cr
&=
([T\sep 0]\,[u,v],[T\sep 0]\,[x,y])_s .
\end{align*}
This shows that $\phat-\qhat$ is an { antilinear} operator
satisfying
\[
(\phat-\qhat)^\star (\phat-\qhat)=[T\sep 0]^2.  
\]

Next calculate
\begin{align*}
(\phat-\qhat)[T\sep 0]^{-1}   
&=
\left(\begin{array}{lr}
       P-Q &0\\
       0 &-(P-Q)
      \end{array}\right)
\left(\begin{array}{lr}
       T^{-1} &0\\
       0 &T^{-1}
      \end{array}\right)\cr
&=
\left(\begin{array}{lr}
       J &0\\
       0 &-J
      \end{array}\right) =
\mhat.
\end{align*}
Because $\mhat$ is an { antilinear}
isometry one concludes that  $$\phat-\qhat=\mhat[T\sep 0]$$ is the polar decomposition
of $\phat-\qhat$.

\endproof

Next, consider the polar decomposition of the operator $\phat+\qhat-\hat\Io$.

\begin{proposition}
One has
\be
\phat+\qhat-\hat\Io=[K\sep 0]\, [C\sep 0].
\label{ortho:revisit:}
\ee
If $C$ is invertible then the above expression is the polar decomposition of the 
operator $\phat+\qhat-\hat\Io$.
\end{proposition}

\beginproof
One has
\begin{align*}
\phat+\qhat-\hat\Io
&=
\left(\begin{array}{lr}
       P+Q-\Io&0\\
       0 &-(P+Q-\Io)
      \end{array}\right)\cr
&=
\left(\begin{array}{lr}
       KC &0\\
       0 &-KC
      \end{array}\right)\cr
&=
\left(\begin{array}{lr}
       KC &0\\
       0 &JKCJ
      \end{array}\right)\cr
&=
[KC\sep 0]\cr
&=
[K\sep 0]\,[C\sep 0].
\end{align*}
Here use is made of Proposition \ref {theorem1} that shows that $J$ and $K$ anti-commute.
In addition $C$ commutes with $J$ because by Proposition \ref{prop:Jsign} $T$ commutes with $J$
and $C=\sqrt{\Io-T^2}$.

Assume now that $C$ is invertible.
For any $[x,y]$ and $[u,v]$ in 
{ ${\mathscr V}\times {\mathscr V}$} one has
\begin{align*}
&
((\phat+\qhat-\hat\Io)[x,y],(\phat+\qhat-\hat\Io)[u,v])_s \cr
&=
([Px,(\Io-P)y]-[(\Io-Q)x,Qy],
[Pu,(\Io-P)v]-[(\Io-Q)u,Qv])_s \cr
&=
([(P+Q-\Io)x,-(P+Q-\Io)y][(P+Q-\Io)u,-(P+Q-\Io)v])_s \cr
&=
s(((P+Q-\Io)x,(P+Q-\Io)u)+s((P+Q-\Io)y,(P+Q-\Io)v)\cr
&
-is((P+Q-\Io)x,J(P+Q-\Io)v)+is(J(P+Q-\Io)y,(P+Q-\Io)u)\cr
&=
s(Cx,Cu)+s(Cy,Cv)
-is(KCx,JKCv)+is(JKCy,KCu)\cr
&=
s(Cx,Cu)+s(Cy,Cv)
+is(Cx,JCv)-is(Cy,JCu)\cr
&=
([Cx,Cy],[Cu,Cv])_s .
\end{align*}
One concludes that $\phat+\qhat-\hat\Io$ is a complex-linear operator
satisfying
\[
(\phat+\qhat-\hat\Io)^\star (\phat+\qhat-\hat\Io)=[C,0]^2.
\]
Because $[K\sep 0]$ is an isometry it now follows that (\ref {ortho:revisit:}) is the polar decomposition
of $\phat+\qhat-\hat\Io$ by uniqueness of the polar decomposition.
\endproof

\subsection{Complexified modular operator}

Next, consider the complexification of the expression for the polar decomposition of $S$.
From Section \ref{complex_antilinear}, both $[S\antisep 0]$ and $[0\antisep S]$ anti-commute with $\ihat$ --- they are { antilinear operators}. Writing out explicitly,
\[
\Splus:=
[S\antisep 0]
=
\left(\begin{array}{lr}
       S &0\\
       0 &-JSJ
      \end{array}\right) =
\left(\begin{array}{lr}
       S &0\\
       0 &-S^\ast
      \end{array}\right)
\]
and
\[
\Splus^\star = [S\antisep 0]^\star
=
\left(\begin{array}{lr}
       S^\ast &0\\
       0 &-JS^\ast J
      \end{array}\right)
=
\left(\begin{array}{lr}
       S^\ast &0\\
       0 &-S
      \end{array}\right),
\]
where $JSJ=S^\ast$ is used. 

Next, define
\[
\Deltaplus:=[\Delta\sep 0]=
\left(\begin{array}{lr}
       \Delta &0\\
       0 &J\Delta J
      \end{array}\right)
=\left(\begin{array}{lr}
       \Delta &0\\
       0 &\Delta^{-1}
      \end{array}\right).
\]
So
\[
\Splus^\star \, \Splus = [S\antisep 0]^\star [S\antisep 0]
=
\left(\begin{array}{lr}
       S^\ast S &0\\
       0 &SS^\ast
      \end{array}\right)
=
\Deltaplus
\]

From
\begin{align*}
\Splus\, \Deltaplus^{\hskip -4pt-1/2}
&=
\left(\begin{array}{lr}
       S &0\\
       0 &-S^*
      \end{array}\right)
\left(\begin{array}{lr}
       \Delta^{-1/2} &0\\
       0 &\Delta^{1/2}
      \end{array}\right)\cr
&=
\left(\begin{array}{lr}
       J &0\\
       0 &-J
      \end{array}\right)\cr
&=
[J\antisep 0]\cr
&=\mhat
\end{align*}
one concludes that
\[
\Splus =\mhat \Deltaplus^{\hskip -4pt1/2}
\]
is the polar decomposition of $\Splus$. That is to say, $\Deltaplus$ is the modular operator on the complex Hilbert space $\mathscr H$
and $\mhat$ is the corresponding modular conjugation operator.

Similarly, $[0\antisep S]$ given by
\[
[0\antisep S]=
\left(\begin{array}{lr}
       0 &SJ\\
       JS &0
      \end{array}\right)
= \left(\begin{array}{lr}
       0 &\Delta^{-1/2}\\
       \Delta^{1/2} &0
      \end{array}\right)
\]
satisfies 
$$ [0\antisep S]^\star [0\antisep S]=\Deltaplus . $$
So it has the polar decomposition 
$$ [0\antisep S]=\hhat \Deltaplus^{\hskip -4pt 1/2}. $$ 

\subsection{The generic pair}

Finally, the following result shows that the antilinear operator
$\Splus$ plays the role of the $S$-operator of Tomita-Takesaki theory 
w.r.t.~the real subspaces ${\mathscr K}_\oplus$ and ${\mathscr L}_\oplus$ defined as
\[
\Kplus={\mathscr K}\oplus{\mathscr K}^\perp
\quad\mbox{ and }\quad
\Lplus={\mathscr L}\oplus{\mathscr L}^\perp.
\]

\begin{proposition}
If $[x,y]$ belongs to $\Kplus$ then $\Splus\,[x,y]=[x,y]$.
If $[x,y]$ belongs to $\Lplus$ then $\Splus\,[x,y]=-[x,y]$.
\end{proposition}
\beginproof
Assume that $[x,y]$ belongs to $\Kplus$. 
This means that $x$ belongs to $\mathscr K$ and $y$ is orthogonal to $\mathscr K$.
The latter implies that $Jy$ belongs to $\mathscr L$.
One has
\[
\Splus\,[x,y]
=
[Sx,-S^*y].
\]
Because $x$ belongs to $\mathscr K$ it satisfies $Sx=x$.
From $JS^*y=SJy$ and $Jy\in{\mathscr L}$ one concludes that $JS^*y=-Jy$
so that $S^*y=-y$.
This proves the former of the two statements.
The proof of the other statement is similar.
\endproof

The above proposition implies that
{ 
\begin{align*}
\Kplus
&=\{[x,y] \in \mathscr V \times \mathscr V:\,\Splus[x,y]=[x,y]\},\cr
\Lplus
&=\{[x,y] \in \mathscr V \times \mathscr V:\,\Splus[x,y]=-[x,y]\}.
\end{align*}
}
This is because $\Splus[x,y]=\pm [x,y]$ is equivalent to $Sx=\pm x$ and $SJy=\mp Jy$, and that $J{\mathscr L}={\mathscr K}^\perp$ and $J{\mathscr K}={\mathscr L}^\perp$.

\begin{proposition}
A vector $[x,y]$ belongs to $\Kplus$ if and only if $\ihat[x,y]$ belongs to $\Lplus$. In particular, if $x$ belongs to $\mathscr K$ then $[x,0]$ belongs to $\Kplus$ and $\ihat [x,0]$
belongs to $\Lplus$.
\end{proposition}

\beginproof
This follows from $\ihat[x,y]=[Jy,-Jx]$ and item (a) of Proposition \ref{prop:Jsign}.
\endproof

\subsection{Example: Larmor precession}
\label{sect:exam}

The example belongs to Quantum Mechanics and is known as Larmor precession of
a quantum spin.
The real vector space $\mathscr V$ 
is taken equal to $M_2(\Ro)$, the space of 2-by-2 matrices with real entries.
Its algebraic structure is neglected in what follows.

Choose a basis $\{\evec_i\}_{i=1}^{4}$ in $\mathscr V$ depending on a single parameter  $r$
and given by
\begin{align*}
&\evec_1=\left(\begin{array}{lr}
       1 &0\\ 0 & 0
      \end{array}\right),
\qquad&
\evec_2=\left(\begin{array}{lr}
       0 &0\\ 0 & 1
      \end{array}\right),\cr
&\evec_3=\left(\begin{array}{lr}
       0 &1-r\\ 1+r & 0
      \end{array}\right),
\qquad&
\evec_4=\left(\begin{array}{lr}
       0 &1-r\\ -(1+r) & 0
      \end{array}\right).
\end{align*}
The basis is used to define the subspaces $\mathscr K$ and $\mathscr L$.
The former is a 3-dimensional space spanned by the vectors $\evec_1$, $\evec_2$, $\evec_3$ while the latter is a 1-dimensional space generated by $\evec_4$:
$$
\mathscr K = \mbox{span} \{ \evec_1, \evec_2, \evec_3 \} , \,\,\,\,\,\,\,\,
\mathscr L = \Ro \evec_4.
$$

The relative geometry of the spaces $\mathscr K$ and $\mathscr L$ depends on
the choice of inner product $s$.
Let the inner product $s$
be given by the trace of a matrix product:
\[
s(x,y)=\Tr (1-r\sigma_3)\, y^\tinyT \,x,
\qquad
x,y\in {\mathscr V},
\]
with $\sigma_3=\evec_1-\evec_2$ and
with $-1<r<1$, $r\not=0$.
Note that the basis $\{\evec_i\}_{i=1}^{4}$ is not orthogonal for this inner product.
The metric tensor $s_{ij}=s(\evec_i,\evec_j)$, is given by
\[
(s_{ij})_{i,j=1}^{4}
=
\left(\begin{array}{lccr}
        1  &0 &0 &0\\
        0 &1  &0 &0\\
        0 &0 &2(1-r^2) &-2r(1-r^2)\\
        0 &0 &-2r(1-r^2) &2(1-r^2)
       \end{array}\right).
\]

The orthogonal projections $P$ and $Q$ clearly satisfy $P\evec_i=\evec_i$ for $i=1,2,3$ and $Q\evec_4=\evec_4$.
One has $P\evec_4=-r\evec_3$ and $Q\evec_1=Q\evec_2=0$, $Q\evec_3=-r\evec_4$.
This implies 
\begin{align*}
(P-Q)\evec_1&=\evec_1,\cr
(P-Q)\evec_2&=\evec_2,\cr
(P-Q)\evec_3&=\evec_3+r\evec_4,\cr
(P-Q)\evec_4&=-r\evec_3-\evec_4.
\end{align*}
So one calculates $T^2=(P-Q)^2$ to obtain
\begin{align*}
T^2 \evec_1 &= \evec_1,\cr
T^2 \evec_2 &= \evec_2,\cr
T^2 \evec_3 &= (1-r^2)\evec_3,\cr
T^2 \evec_4 &= (1-r^2)\evec_4.
\end{align*}

Introduce an orthogonal basis $\{\mvec_i\}_{i=1}^{4}$ defined by
$\mvec_1=\evec_1$, $\mvec_2=\evec_2$,  $\mvec_3=\evec_3+\evec_4$, $\mvec_4=\evec_3-\evec_4$.
It diagonalizes $T^2$, and hence its square-root $T$, with
$T$ given by
\begin{align*}
T\mvec_1&=\mvec_1,\cr
T\mvec_2&=\mvec_2,\cr
T\mvec_3&=(1-r^2)^{1/2}\mvec_3,\cr
T\mvec_4&=(1-r^2)^{1/2}\mvec_4.
\end{align*}
The isometry $J$ follows now from $J=(P-Q)T^{-1}$, explicitly given in the original basis as
\begin{align*}
J\evec_1&=\evec_1, \cr
J\evec_2&=\evec_2, \cr
J\evec_3&=(1-r^2)^{-1/2}(\evec_3+r\evec_4),\cr
J\evec_4&=-(1-r^2)^{-1/2}(r\evec_3+\evec_4).
\end{align*}
One readily verifies $J^2=\Io$.
Note that
\[
s(J\evec_3,\evec_3)
=(1-r^2)^{-1/2}s(\evec_3+r\evec_4,\evec_3)
=2(1-r^2)^{3/2}
>0
\]
and
\[
s(J\evec_4,\evec_4)
=-(1-r^2)^{-1/2}s(r\evec_3+\evec_4,\evec_4)
=-2(1-r^2)^{3/2}
<0 ,
\]
both as expected from Proposition \ref {prop:Jsign}. 

The operator $S$ is defined by $S\evec_i=\evec_i$ for $i=1,2,3$ and $S\evec_4=-\evec_4$.
The square root of the modular operator $\Delta$ then follows from
$\Delta^{1/2}=JS$, or explicitly
\begin{align*}
\Delta^{1/2}\evec_1&=\evec_1, \cr
\Delta^{1/2}\evec_2&=\evec_2, \cr
\Delta^{1/2}\evec_3&=(1-r^2)^{-1/2}(\evec_3+r\evec_4),\cr
\Delta^{1/2}\evec_4&=(1-r^2)^{-1/2}(r\evec_3+\evec_4).
\end{align*}
The operator $\Delta^{1/2}$ is diagonal in the orthogonal basis $\{\mvec_i\}_{i=1}^4$
with eigenvalues $1$, $1$, $(1-r^2)^{-1/2}(1+r) \equiv ((1+r)/(1-r))^{1/2}$, $(1-r^2)^{-1/2}(1-r) \equiv ((1-r)/(1+r))^{1/2}$.

The Hamiltonian $H=\log\Delta$ satisfies $H\mvec_1=H\mvec_2=0$ and
\be
H\mvec_3=\omega \mvec_3
\quad\mbox{ and }\quad
H\mvec_4=-\omega \mvec_4
\label{ex:ham}
\ee
with
\[
\omega=\frac 14\log\frac{1+r}{1-r}.
\]
This implies $H\evec_1=H\evec_2=0$ and
\be
H\evec_3=\omega \evec_4
\quad\mbox{ and }\quad
H\evec_4=\omega \evec_3.
\label{ex:ham2}
\ee

For any element $\yvec$ in $\mathscr K$ with the form
\[
\yvec=\sum_{i=1}^3a_i\evec_i
\]
and any element $\zvec$ in $\mathscr L$ of the form
\[
\zvec=a_4\evec_4, 
\]
one obtains, from (\ref {ex:ham2}) that
\begin{align*}
\dot \yvec
&= \,\, H\zvec \,\, = \,\, a_4\omega \evec_3,\cr
\dot \zvec &= \,\, -H\yvec \,\, = \,\,
-a_3 \omega \evec_4.
\end{align*}
This leads to  the set of equations $\dot a_1=\dot a_2=0$ and
\begin{align*}
\dot a_3&=\omega a_4,\cr
\dot a_4&=-\omega a_3.
\end{align*}
Hence $a_1$ and $a_2$ are constant and $a_3,a_4$ describe harmonic oscillations
\begin{align*}
a_3(t)&=a_3(0)\,\cos(\omega t)+a_4(0)\,\sin(\omega t),\cr
a_4(t)&=-a_3(0)\,\sin(\omega t)+a_4(0)\,\cos(\omega t).
\end{align*}

To make the connection with Larmor precession,
introduce the notations
\[
\sigma_+=
\left(\begin{array}{lr}
       0 &1\\ 0 & 0
      \end{array}\right),
\quad
\sigma_-=
\left(\begin{array}{lr}
       0 &0\\ 1 & 0
      \end{array}\right).
\]
They satisfy
\[
\sigma_+=\frac 12\frac{1}{1-r}\mvec_3
\quad\mbox{ and }\quad
\sigma_-=\frac 12\frac{1}{1+r}\mvec_4
\]
and
\[
J\sigma_+=\frac{1+r}{\sqrt{1-r^2}}\sigma_-
\quad\mbox{ and }\quad
J\sigma_-=\frac{1-r}{\sqrt{1-r^2}}\sigma_+.
\]
This implies
\[
H\sigma_\pm=\pm \omega\sigma_\pm
\]
and
\[
HJ\sigma_\pm=\mp\omega J\sigma_\pm .
\]

The time evolution of these matrices in the complexified Hilbert space $\mathscr H$
is given by 
\[
\Deltaplusit[\sigma_3,0]=[\sigma_3,0]
\]
and
\begin{align*}
\Deltaplusit[\sigma_\pm, 0]
&=
\left(\cos(t\Hplus)+\ihat\sin(t\Hplus)\right)\,[\sigma_\pm, 0]\cr
&=
[\cos(tH)\sigma_\pm,0]+[0,\sin(tH)J\sigma_\pm]\cr
&=
\cos (\omega t)[\sigma_\pm,0]\mp\sin(\omega t)[0,J\sigma_\pm]\cr
&=
\cos (\omega t) [\sigma_\pm,0]\pm \ihat \sin(\omega t)[\sigma_\pm,0]\cr
&=
\evec^{\pm \ihat \omega t}[\sigma_\pm, 0]
\end{align*}      
with
\[
\Hplus=\log\Deltaplus=
\left(\begin{array}{lr}
       H &0\\
       0 &-H
      \end{array}\right).
\]
Here, use is made of $H\sigma_\pm=JHJ\sigma_\pm=\omega\sigma_\pm$.
The above equations are known as Larmor precession of a spin
in the Heisenberg picture of quantum mechanics.

\section{Conclusion and Discussions}

This work considers a pair of subspaces $\mathscr K$ and $\mathscr L$ of a
finite-dimensional real Hilbert space $\mathscr V$.
The only assumptions are that the intersection of these two subspaces
is trivial and that the sum is the total space $\mathscr V$.
The corresponding orthogonal projection operators, denoted $P$ and $Q$, do not commute in general.
Following Rieffel and van Daele  \cite{RvD77}, $P$ and $Q$ determine two operators
which are the analogues of the modular conjugation operator $J$ and the modular operator $\Delta$
of Tomita-Takesaki theory \cite{TM70}. The operator $S$ of Tomita-Takesaki theory
has an obvious definition in terms of the two subspaces $\mathscr K$ and $\mathscr L$, see (\ref {modop:Sdef}).
Its properties are proved in Proposition \ref{prop:props}.

The operator $J$ is characterised in Proposition \ref{prop:Jsign} as the unique self-adjoinnt orthogonal
operator with certain specific properties w.r.t.~the subspaces $\mathscr K$ and $\mathscr L$.
In Section \ref{sect:complex} it is used to add a complex structure to the product space
{
${\mathscr V}\times{\mathscr V}$.}
A representation of Klein's group of order 4 is
formed by the operators $\Io,-\Io,J,-J$. It combines with
one of the two cyclic groups of order 4 (involving $\ihat$ or $\dhat$, respectively)
to form a group of order 16 which is a central extension of the group of split quaternions.
Either of these  two elements $\ihat$ or $\dhat$ can be chosen to complexify the Hilbert space
{
${\mathscr V}\times{\mathscr V}$.
}

The proofs of some of the results of \cite{RvD77} are reproduced here for the sake of readability
of the paper. What is new is our focus on the case that the pair of subspaces
is {\it not} in generic position in the sense of Halmos \cite{HPR69}. 
To this purpose we consider
the analogue of the isometry $J$ obtained when replacing the subspace $\mathscr L$ by its orthogonal complement ${\mathscr L}^\perp$, namely, the $K$-operator. $K$ is again an isometry when the two subspaces are in generic position.
If the pair of subspaces is not in generic position, then the operator $K$ is only a partial
isometry ---  its null space is denoted $\mathscr S$. 
It is the {\it symmetric subspace} of Section \ref {sect:symsub}.

Proposition \ref {theorem1} shows that the partial isometries $J$ and $K$ anti-commute.
This is a new and unexpected result.

Proposition \ref {symm:prop:props} characterizes the subspace  $\mathscr S$.
If $\mathscr S$ is trivial then the pair of subspaces ${\mathscr K}$ and ${\mathscr L}$
is in generic position. If it equals the total space $\mathscr V$ then the two subspaces are
orthogonal to each other. So introducing the $K$-operator allows us to investigate
the full range of possibilities of the two subspaces ${\mathscr K}$ and ${\mathscr L}$
which correspond to a pair of quite arbitrary projectors $P$ and $Q$
on a finite-dimensional real vector space ${\mathscr V}$.

The notations introduced in Section \ref{sect:complex} require some justification.
Vectors in the product space 
{
${\mathscr V}\times{\mathscr V}$
} are denoted $[x, y]$
rather than $(x,  y)^\tinyT$ for the sake of clarity. Alternatives are $x\oplus y$ and
$x+iy$. The latter is avoided because it leads to confusion about the vectors being considered
as members of a real or of a complex Hilbert space. For the same reason operators
on the product space are denoted $[A\sep B]$ instead of $A+iB$ if they are complex-linear
and $[A\antisep B]$ if they are conjugate-linear.

Two examples are given. The example of Section \ref {sect:eg:eucl}
considers a pair of subspaces with unequal dimensions.
This implies that the pair is not in generic position.
The example of Section \ref {sect:exam} treats a two-dimensional model of quantum mechanics.

 In a companion paper \cite{NJ25} the complexification procedure is applied to
 Riemannian manifolds the tangent spaces of which can be decomposed into two
 subspaces $\mathscr K$ and $\mathscr L$. The Theory of Linear Response, as found in \cite{NVW75}
 is generalized and a new Fluctuation-Dissipation Theorem \cite{KR66} is derived.

An obvious application of the present work is the study of actions of the real line in a real Hilbert space
with a Hamiltonian which is the logarithm of the modular operator $\Delta$.
Such an action defines a Hamiltonian flow \cite{MR99}
and is expected to satisfy a form of Kubo-Martin-Schwinger (KMS) condition \cite{HHW67}.
In Statistical Physics, the KMS condition \cite{HHW67,RD69,BR79,PJ84,PD08} 
characterizes the quantum Gibbs states and the quantum-mechanical time evolution.


Complications arise when replacing the complexification operator of the pres\-ent study by an almost complex operator $I$ acting on the tangent bundle and satisfying $I^2=-\Io$, which may not be integrable. Its integration gives rise to holomorphic coordinates on the manifold $\Mo$, but
its dimensionality needs to be even to begin with. 
The integrability condition is a topological one, governed by the vanishing of the so-called Nijenhuis tensor. 
In the present work, the vector space ${\mathscr V}$ on which the pair of subspaces is defined
does not need to be of even dimension. The transition from $P,Q$ projections in ${\mathscr V}$
to $\phat, \qhat$ projections in 
{ ${\mathscr V}\times{\mathscr V}$} requires a doubling of the dimensionality that is not immediately obvious when applying the current apparatus to the manifold context.

Two possible solutions come to mind: 
1) It is possible to consider $T_p \Mo \oplus T^\ast_p \Mo $ for $p\in \Mo$. 
As a generalized complex structure, this has a natural symplectic structure, and
much of the apparatus of the present paper can be moved over.
2) It is also possible to consider the split-quaternion structure directly by exploring the canonical split of $T(T \Mo)$ as induced from an affine connection and a Riemannian metric on $\Mo$. Note that the tangent manifold $T\Mo$ is always even-dimensional, and its tangent vectors (which live in the iterated tangent space $T(T\Mo)$) have a natural split in the horizontal (fiber) and vertical directions, controlled by the affine connection on $\Mo$. Furthermore, the Riemannian metric on $\Mo$ can be lifted to a Sasaki metric on $T\Mo$, generating a split-quaternion structure.
It is known that a flat connection on $\Mo$ can induce an integrable complex structure on $T\Mo$ \cite{Dombrowski}, whereas a pair of dually flat connections (with respect to a certain Hessian metric) will induce a pair of K\"ahler structures on $T\Mo$ \cite{Shima, ZhangKhan}. This latter scenario, for which the current apparatus is readily applicable, is more interesting to information geometers, since dually flat structures on the base manifold $\Mo$ arise abundantly from exponential families of probability density functions \cite{AN00}.

\nocite{*}
\bibliography{kms24v5}{}

\begin{thebibliography}{10}

\bibitem{AN00}
S.~Amari and H.~Nagaoka.
\newblock {\em Methods of Information Geometry}.
\newblock Oxford University Press, 2000.

\bibitem{BR79}
O.~Bratteli and D.W. Robinson.
\newblock {\em Operator Algebras and Quantum Statistical Mechanics. Vol I, II.
  2nd edition.}
\newblock Springer, New York, Berlin, 1979.

\bibitem{dG06}
M.~de~Gosson.
\newblock {\em Symplectic Geometry and Quantum Mechanics}.
\newblock Birkh{\"a}user, 2006.

\bibitem{Dombrowski}
P.~Dombrowski.
\newblock On the geometry of the tangent bundle.
\newblock {\em Journal f{\"ur} die reine und angewandte Mathematik},
  210:73--88, 1962.

\bibitem{HHW67}
R.~Haag, N.~M. Hugenholtz, and M.~Winnink.
\newblock On the equilibrium states in quantum statistical mechanics.
\newblock {\em Commun. math. Phys.}, 5:215--236, 1967.

\bibitem{HPR69}
P.~R. Halmos.
\newblock Two subspaces.
\newblock {\em Trans. Am. Math. Soc.}, 144:381--389, 1969.

\bibitem{KT66}
T.~Kato.
\newblock {\em Perturbation theory for linear operators}.
\newblock Springer-Verlag, Berlin-Heidelberg-New York, 1966.

\bibitem{KR66}
R.~Kubo.
\newblock The fluctuation-dissipation theorem.
\newblock {\em Rep. Progr. Phys.}, 29:255--284, 1966.

\bibitem{MR99}
J.~E. Marsden and T.~S. Ratiu.
\newblock {\em Introduction to mechanics and symmetry}.
\newblock Springer, 1999.

\bibitem{NJ25}
J.~Naudts.
\newblock A complex structure for two-typed tangent spaces.
\newblock {\em Entropy}, 27:117, 2025.

\bibitem{NVW75}
J.~Naudts, A.~Verbeure, and R.~Weder.
\newblock Linear response theory and the {KMS} condition.
\newblock {\em Commun. Math. Phys.}, 44:87--99, 1975.

\bibitem{PD08}
D.~Petz.
\newblock {\em Quantum Information Theory and Quantum Statistics}.
\newblock Springer-Verlag, Berlin, 2008.

\bibitem{PJ84}
J.~Pul{\`e}.
\newblock A unified approach to classical and quantum {K.M.S.} theory.
\newblock {\em Rep. Math. Phys.}, 20:75--81, 1984.

\bibitem{RvD77}
M.~A. Rieffel and A.~van Daele.
\newblock A bounded operator approach to {Tomita-Takesaki} theory.
\newblock {\em Pac. J. Math.}, 69:187--221, 1977.

\bibitem{RW91}
W.~Rudin.
\newblock {\em Functional Analysis. 2nd Ed.}
\newblock Mc Graw-Hill, 1991.

\bibitem{RD69}
D.~Ruelle.
\newblock {\em Statistical mechanics, Rigorous results}.
\newblock W.A. Benjamin, Inc., New York, 1969.

\bibitem{Shima}
H.~Shima.
\newblock {\em The Geometry of {Hessian} Structures}.
\newblock World Scientific, 2007.

\bibitem{TM70}
M.~Takesaki.
\newblock {\em Tomita's theory of modular Hilbert algebras and its
  applications. Lecture Notes in Mathematics 128}.
\newblock Springer-Verlag, Berlin, 1970.

\bibitem{ZhangKhan}
J.~Zhang and G.~Khan.
\newblock Statistical mirror symmetry.
\newblock {\em Differential Geometry and its Applications}, 73:101678, 2020.

\end{thebibliography}
\bibliographystyle{plain}


\end{document}